\documentclass{sig-alternate}

\usepackage{booktabs} % nicer tables
\usepackage[skip=5pt]{caption} % allow captions without figure/table labels
\usepackage{subfigure} % allow sub captions in figures
\usepackage{ccicons} % creative commons icons
\usepackage{enumitem} % control items, lists, descriptions
\usepackage{todonotes} % todo notes
\usepackage{flushend} % balance columns on last page
\usepackage[numbers,sort&compress]{natbib} % citations
\usepackage[hang,flushmargin]{footmisc} % control footnotes
\usepackage{xcolor} % control color use
\usepackage[hyphens]{url} % handle long urls
\usepackage[bookmarks, pdftex, colorlinks=false, pdfpagelabels=false]{hyperref} % clickable references; note: general rule seems to be to always load hyperref last

\newfont{\mycrnotice}{ptmr8t at 7pt}
\newfont{\myconfname}{ptmri8t at 7pt}
\let\crnotice\mycrnotice%
\let\confname\myconfname%
\permission{Publication rights licensed to ACM. ACM acknowledges that this contribution was authored or co-authored by an employee, contractor or affiliate of the United States government. As such, the United States Government retains a nonexclusive, royalty-free right to publish or reproduce this article, or to allow others to do so, for Government purposes only.}
\copyrightetc{Copyright is held by the owner/author(s)}
\conferenceinfo{Communications of the ACM}{0001-0782/2016/2 \$15.00\\
http://dx.doi.org/10.1145/2812802}
\begin{document}

% Spacing
\frenchspacing

% Title
\title{YFCC100M: The New Data in Multimedia Research}

% Authors
\numberofauthors{8}
\author{%
\alignauthor Bart Thomee \\
\affaddr{Yahoo Labs} \\
\affaddr{\mbox{San Francisco, CA, USA}} \\
\email{\mbox{bthomee@yahoo-inc.com}}
\alignauthor David A.~Shamma \\
\affaddr{Yahoo Labs} \\
\affaddr{\mbox{San Francisco, CA, USA}} \\
\email{\mbox{aymans@acm.org}}
\alignauthor Gerald Friedland \\
\affaddr{ICSI} \\
\affaddr{\mbox{Berkeley, CA, USA}} \\
\email{\mbox{fractor@icsi.berkeley.edu}}
\and
\alignauthor Benjamin Elizalde\thanks{\enspace This work was done while Benjamin Elizalde was at ICSI.} \\
\affaddr{CMU} \\
\affaddr{\mbox{Mountain View, CA, USA}} \\
\email{\mbox{bmartin1@andrew.cmu.edu}}
\alignauthor Karl Ni\thanks{\enspace This work was done while Karl Ni was at LLNL.} \\
\affaddr{In-Q-Tel} \\
\affaddr{\mbox{Menlo Park, CA, USA}} \\
\email{\mbox{kni@iqt.org}}
\alignauthor Douglas Poland \\
\affaddr{LLNL} \\
\affaddr{\mbox{Livermore, CA, USA}} \\
\email{\mbox{poland1@llnl.gov}}
\and
\alignauthor Damian Borth\thanks{\enspace This work was done while Damian Borth was at ICSI.} \\
\affaddr{DFKI} \\
\affaddr{\mbox{Kaiserslautern, Germany}} \\
\email{\mbox{damian.borth@dfki.de}}
\alignauthor Li-Jia Li\thanks{\enspace This work was done while Li-Jia Li was at Yahoo Labs.} \\
\affaddr{Snapchat} \\
\affaddr{\mbox{Venice, CA, USA}} \\
\email{\mbox{lijiali.vision@gmail.com}}
}
\maketitle

% !TEX root = CACM2015_YFCC100M.tex

\noindent The photograph and our understanding of photography is ever changing and has transitioned from a world of unprocessed rolls of C-41 sitting in a fridge 50 years ago to sharing photos on the 1.5" screen of a point and shoot camera 10 years back. And today the photograph is again something different. The way we take photos is fundamentally different. We can view, share, and interact with photos on the device they were taken on. We can edit, tag, or ``filter'' photos directly on the camera at the same time the photo is being taken. Photos can be automatically pushed to various online sharing services, and the distinction between photos and videos has lessened. Beyond this, and more importantly, there are now lots of them. To Facebook alone more than 250 billion photos have been uploaded and on average it receives over 350 million new photos every day~\cite{Facebook:2013ab}, while YouTube reports that 300 hours of video are uploaded every minute~\cite{YouTube:2015ab}. A back of the envelope estimation reports 10\% of all photos in the world were taken in the last 12 months, and that was calculated already more than three years ago~\cite{good:photos:url}.

Today, a large number of the digital media objects that are shared have been uploaded to services like Flickr or Instagram, which along with their metadata and their social ecosystem form a vibrant environment for finding solutions to many research questions at scale. Photos and videos provide a wealth of information about the universe, covering entertainment, travel, personal records, and various other aspects of life in general as it was when they were taken. Considered collectively, they represent knowledge that goes beyond what is captured in any individual snapshot and provide information on trends, evidence of phenomena or events, social context, and societal dynamics. Consequently, collections of media are useful for qualitative and quantitative empirical research in many domains. However, scientific endeavors in fields like social computing and computer vision have generally relied on independently collected multimedia datasets, which complicates research growth and synergy. There is the need for a more substantial dataset for researchers, engineers, and scientists around the globe.

To meet the call for scale, openness, and diversity in academic datasets, we take the opportunity in this article to present a new multimedia dataset containing 100 million media objects and explain the rationale behind its creation. We discuss the implications it has for science, research, engineering and development, and demonstrate its usefulness towards tackling a broad range of problems in various domains. With the release of our dataset comes the opportunity to advance research, giving rise to new challenges and solving existing ones.

%%% Local Variables: 
%%% mode: latex
%%% TeX-master: "TMM2015_YFCC100M"
%%% End: 

% !TEX root = CACM2015_YFCC100M.tex

\vspace{-3pt}
\section{Sharing Datasets}
\label{section:sharing}
\noindent Datasets are critical for research and exploration~\cite{Renear:2010tl} as, rather obviously, data is required for performing experiments, validating hypotheses, analyzing designs, and building applications. Over the years a plurality of multimedia datasets have been put together for research and development. In Table~\ref{table:highlight} we present a summary of the most popular datasets at present. Most of them, however, actually cannot truly be called multimedia datasets due to only containing a single type of media, rather than a mixture of modalities, such as photos, videos, and audio. Datasets range from the one-off instances that have been exclusively created for supporting the work presented in a single paper or demo (`short-term datasets') to those that have been created with multiple related or separate endeavors in mind (`long-term datasets'). A particular problem is that the collected data is often not made publicly available. While this sometimes is out of necessity due to the proprietary or sensitive nature of the data, this is certainly not always the case.

\begin{table*}[t]
\centering
\caption{Popular multimedia datasets used over the years by the research community. When various versions of a particular collection were available, we generally only included the most recent one. PASCAL, TRECVID, MediaEval and ImageCLEF are/were yearly recurring benchmarks that consist of one or more challenges of which each has its own dataset; in the table we report the total number of media objects aggregated over all datasets part of the most recent edition of each benchmark.}
\label{table:highlight}
\vspace{-4pt}
\includegraphics[trim=8 8 8 8, clip=true, width=\textwidth]{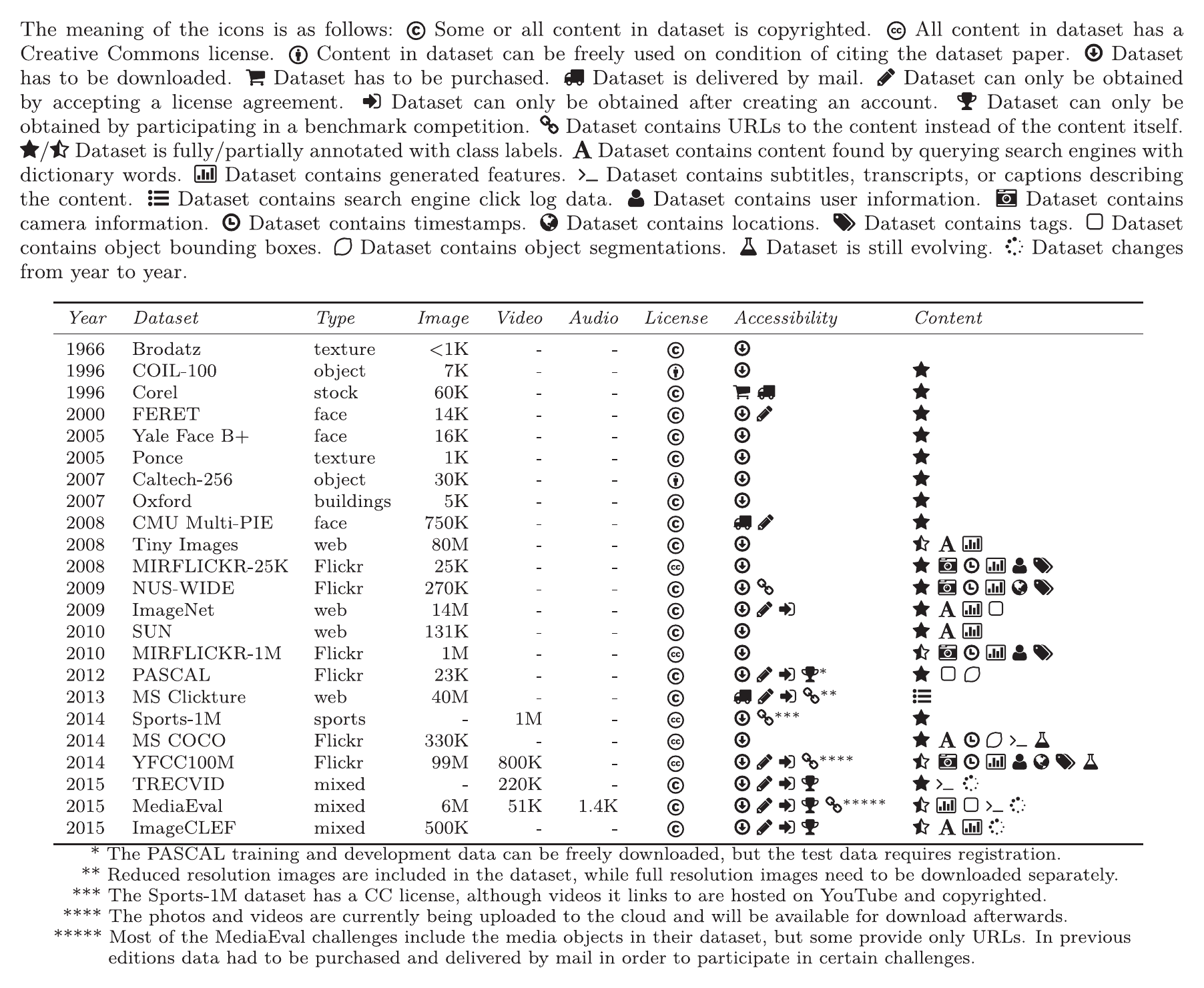}
\vspace{-12pt}
\end{table*}

The question of sharing data for replication and growth has arisen several times in the past 30 years alone~\cite{Fienberg:1985sh,Swan:2008ab,Borgman:2012en} and has been brought into discussion in ACM's SIGCHI~\cite{Wilson:2014:RWI:2559206.2559233}. This ``sharing discussion'' reveals many of the underlying complexities with sharing, both with regards to the data (e.g.\ what exactly is considered data) and to the sharing point of view (e.g.\ incentives and disincentives for doing so). For example, one might be reluctant to share data freely, as it has a value from the often substantial amount of time, effort, and money that was invested in collecting it.

Another barrier to sharing arises when data is harvested for research under a general umbrella of ``academic fair use'' without regards towards its licensing terms. Beyond the legal corporate issues, this clearly may violate the copyright of the owner of the data that, in many User-Generated Content (UGC) sites like Flickr, remains with the creator. The Creative Commons (CC), a nonprofit organization that was founded in 2001, seeks to build a rich public domain of \textit{``some rights reserved''} media, sometimes referred to as the copyleft movement. Their licenses allow media owners to communicate how they would like their media to be rights reserved. For example, an owner can indicate that a photo may be used only for non-commercial purposes or whether someone is allowed to remix it or turn it into a collage. Depending on how the licensing options are chosen, CC licenses can be applied that are more restrictive (e.g.\ \texttt{CC Attribution-NonCommercial-NoDerivs} or \texttt{CC-BY-NC-ND} license) or less restrictive (e.g.\ \texttt{CC Attribution-ShareAlike} or \texttt{CC-BY-SA}) in nature. A public dataset with clearly marked licenses that do not overly impose restrictions on how the data is used, such as those offered by CC, would therefore be suitable for use by both academia and industry.

We underscore the importance of sharing---perhaps even its principal argument---is that it ensures \textit{data equality} for research. While the availability of data alone may not necessarily be sufficient for the exact reproduction of scientific results (since the original experimental conditions also would need to be replicated as closely as possible, which may not always be possible), research should start with publicly sharable and legally usable data that is flexible and rich enough to promote new advancements, rather than with data that only serves as a one time collection for a specific task and that cannot be shared. Shared datasets can play a singular role in achieving research growth and in facilitating synergy within the research community that is otherwise difficult to attain.

%%% Local Variables: 
%%% mode: latex
%%% TeX-master: "TMM2015_YFCC100M"
%%% End: 

% !TEX root = CACM2015_YFCC100M.tex

\section{The YFCC100M Dataset} 
\label{section:dataset}
\noindent We created the Yahoo Flickr Creative Commons 100 Million Dataset\footnote{The dataset is available at \url{https://bit.ly/yfcc100md}.} (YFCC100M) as part of the Yahoo Webscope program. This dataset is the largest public multimedia collection that has ever been released, comprising a total of 100 million media objects, of which approximately 99.2 million are photos and 0.8 million are videos, all of which have been uploaded to Flickr between 2004 and 2014 and published under a CC commercial or non-commercial license. The dataset is currently distributed via Amazon AWS as a 12.5GB compressed archive containing only metadata. However, as is the case with many datasets, the YFCC100M is constantly evolving, and over time we will release various expansion packs containing data not yet present in the collection. For instance, several visual and aural features extracted from the data have already been made available\footnote{The features are available at \url{https://bit.ly/yfcc100mf}.}. The actual photos and videos are further currently being uploaded to the cloud to ensure the dataset will remain accessible and intact for years to come; like the metadata, the photo and video data can then be mounted as a read-only network drive. Our dataset overcomes many of the issues affecting existing datasets, for instance in terms of \textit{modalities}, \textit{metadata}, \textit{licensing}, and principally \textit{volume}---we will discuss the strengths and limitations of our collection in more detail below.

\subsection{Metadata}
\noindent Each media object included in the dataset is represented by its metadata in the form its Flickr identifier, the user that created it, the camera that took it, the time at which it was taken and when it was uploaded, the location where it was taken (if available), and the CC license it was published under. In addition, the title, description and tags are also available, as well as direct links to its page and its content on Flickr. Social features, comments, favorites, and followers/following data are not included in the dataset as, by their nature, these change on a day to day basis. They can, however, be easily obtained by querying the Flickr API\footnote{\url{https://www.flickr.com/services/api/}}, and over time we will make snapshots of such social features available for download. We are further working towards releasing the Exif metadata of the photos and videos as an expansion pack.

\begin{description}[leftmargin=0pt,itemsep=4pt]
\item[Tags.] There are 68,552,616 photos and 418,507 videos in the dataset that have been annotated with user tags (or keywords). The tags make for a rich and diverse set of entities related to people (\texttt{baby}, \texttt{family}), animals (\texttt{cat}, \texttt{dog}), locations (\texttt{park}, \texttt{beach}), travel (\texttt{nature}, \texttt{city}), to name just the top few.
A total of 3,343,487 photos and 7,281 videos carry machine tags (labels that have been automatically added by a camera, computer, application, or some other automated system).
\item[Timespan.] Although our dataset contains media \emph{uploaded} between the inception of Flickr in 2004 and the creation of the dataset in 2014, the actual time span during which they were \emph{captured} is much longer. Some scans of books and newspapers have even been backdated to the early 19th century when they were originally published. However, we note that camera clocks are not always set to the correct date and time, and some photos and videos even erroneously report they were captured in the distant past or in the future. Figure~\ref{figure:timestamps} plots the moments of capture and upload of photos and videos during the period 2000-2014, which accounts for 99.6\% of the media objects in the dataset.

\begin{figure}[t!]
\vspace{6pt}
\centering
\includegraphics[trim=0 2 0 0, clip=true, width=\columnwidth]{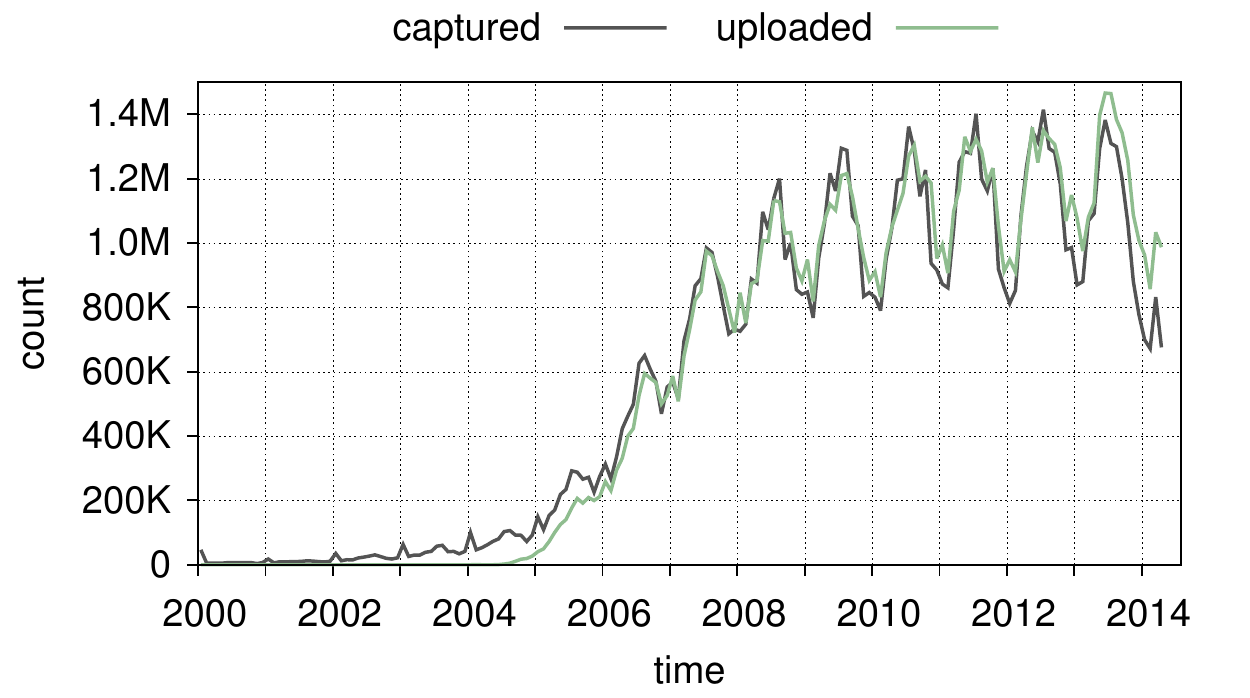}
\caption{Number of captured and uploaded media objects per month in the YFCC100M dataset during the period 2000-2014. We can see the number of uploads closely following the number of captures, with more recently uploads even exceeding captures as older media also is uploaded.}
\label{figure:timestamps}
\end{figure}

\item[Locations.] There are 48,366,323 photos and 103,506 videos in the dataset that have been annotated with a geographic coordinate, either manually by the user or automatically via GPS. The cities in which more than 10,000 unique users captured media are London, Paris, Tokyo, New York, San Francisco, and Hong Kong. Overall, the dataset spans 249 different territories (countries, islands, etc.) in the world, and also includes photos and videos taken in international waters and airspace, see Figure~\ref{figure:locations}.

\begin{figure}[t!]
\centering
\includegraphics[trim=0 70 0 70, clip=true, width=0.9\columnwidth]{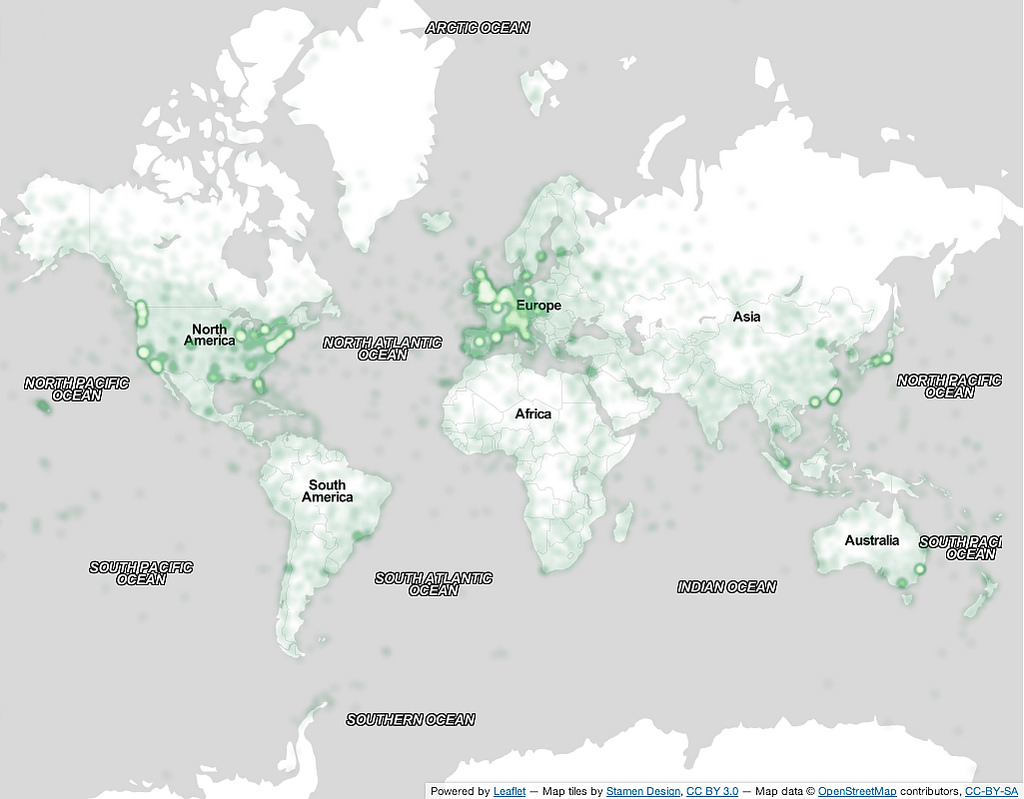}
\caption{Global coverage of a sample of one million photos from the YFCC100M dataset. \textit{One Million Creative Commons Geo-tagged Photos} by David Shamma \ccbynd~\url{https://flic.kr/p/o1Ao2o}.}
\label{figure:locations}
\vspace{-6pt}
\end{figure}

\item[Cameras.] Table~\ref{table:cameras} shows that the top 25 cameras used in the dataset are overwhelmingly digital single lens reflex (DSLR) models with the exception of the Apple iPhone. Considering that the most popular cameras in the Flickr community at the moment primarily consist of various iPhone models\footnote{\url{https://www.flickr.com/cameras/}}, this bias in our data is likely due to CC licenses attracting a certain subcommunity of photographers that differs from the overall Flickr user base.
\item[Licenses.] The licenses themselves vary by CC type with approximately 31.8\% of the dataset marked appropriate for commercial use and 17.3\% having been assigned the most liberal license that only requires the photographer that took the photo to be attributed, see Table~\ref{table:licenses}.
\end{description}

\begin{table}
\centering
\caption{Top 25 cameras and photo counts in the YFCC100M dataset. We have merged the entries for the Canon models that have different names in the European (e.g.\ EOS 650D), American (e.g.\ EOS Rebel T4i) and Asian (e.g.\ EOS Kiss X6i) markets.}
\begin{tabular}{l l r}
\toprule \textit{Make} & \textit{Model} & \textit{Photos} \\
\midrule
Canon & EOS 400D & \texttt{2,539,571} \\ % canon+eos+400d+digital: 1211370, canon+eos+digital+rebel+xti: 1232934, canon+eos+kiss+digital+x: 95267
Canon & EOS 350D & \texttt{2,140,722} \\ % canon+eos+350d+digital: 859475, canon+eos+digital+rebel+xt: 1207515, canon+eos+kiss+digital+n: 73732
Nikon & D90 & \texttt{1,998,637} \\ % nikon+corporation+nikon+d90: 1998637
Canon & EOS 5D Mark II & \texttt{1,896,219} \\ % canon+eos+5d+mark+ii: 1896219
Nikon & D80 & \texttt{1,719,045} \\ % nikon+corporation+nikon+d80: 1719045
Canon & EOS 7D & \texttt{1,526,158} \\ % canon+eos+7d: 1526158
Canon & EOS 450D & \texttt{1,509,334} \\ % canon+eos+450d: 734838, canon+eos+digital+rebel+xsi: 774566, canon+eos+kiss+x2: 59650
Nikon & D40 & \texttt{1,358,791} \\ % nikon+corporation+nikon+d40: 1358791
Canon & EOS 40D & \texttt{1,334,891} \\ % canon+eos+40d: 1334891
Canon & EOS 550D & \texttt{1,175,229} \\ % canon+eos+550d: 476754, canon+eos+rebel+t2i: 640555, canon+eos+kiss+x4: 57920
Nikon & D7000 & \texttt{1,068,591} \\ % nikon+corporation+nikon+d7000: 1068591
Nikon & D300 & \texttt{1,053,745} \\ % nikon+corporation+nikon+d300: 1053745
Nikon & D50 & \texttt{1,032,019} \\ % nikon+corporation+nikon+d50: 1032019
Canon & EOS 500D & \texttt{1,031,044} \\ % canon+eos+500d: 441809, canon+eos+rebel+t1i: 506556, canon+eos+kiss+x3: 82679
Nikon & D700 & \texttt{942,806} \\ % nikon+corporation+nikon+d700: 942806
Apple & iPhone 4 & \texttt{922,675} \\ % apple+iphone+4: 922675
Nikon & D200 & \texttt{919,688} \\ % nikon+corporation+nikon+d200: 919688
Canon & EOS 20D & \texttt{843,133} \\ % canon+eos+20d: 843133
Canon & EOS 50D & \texttt{831,570} \\ % canon+eos+50d: 831570
Canon & EOS 30D & \texttt{820,838} \\ % canon+eos+30d: 820838
Canon & EOS 60D & \texttt{772,700} \\ % canon+eos+60d: 772700
Apple & iPhone 4S & \texttt{761,231} \\ % apple+iphone+4s: 761231
Apple & iPhone & \texttt{743,735} \\ % apple+iphone: 743735
Nikon & D70 & \texttt{742,591} \\ % nikon+corporation+nikon+d70: 742591
Canon & EOS 5D & \texttt{699,381} \\ % canon+eos+5d: 699381
\bottomrule
\end{tabular}
\label{table:cameras}
\vspace{-6pt}
\end{table}

\begin{table}[t]
\centering
\caption{A breakdown of the 100 million photos and videos by their kind of \ccLogo~Creative Commons license, as \ccAttribution~By Attribution, \ccNoDerivatives~No Derivatives, \ccShareAlike~Share Alike, and \ccNonCommercial~Non-Commercial.}
\begin{tabular}{l r r}
\toprule
\textit{License} & \textit{Photos} & \textit{Videos} \\
\midrule
\ccby & \texttt{17,210,144} & \texttt{137,503} \\
\ccbysa & \texttt{9,408,154} & \texttt{72,116} \\
\ccbynd & \texttt{4,910,766} & \texttt{37,542} \\
\ccbync & \texttt{12,674,885} & \texttt{102,288} \\
\ccbyncsa & \texttt{28,776,835} & \texttt{235,319} \\
\ccbyncnd & \texttt{26,225,780} & \texttt{208,668} \\
\midrule Total & \texttt{99,206,564} & \texttt{793,436}\\
\bottomrule
\end{tabular}
\label{table:licenses}
\end{table}

\subsection{Content}
\begin{figure*}
\centering
\subfigure[\textit{IMG\_9793: Streetcar (Toronto Transit)} by Andy Nystrom \ccbyncnd~\url{https://flic.kr/p/jciMdz}.]{\includegraphics[width=0.4\textwidth]{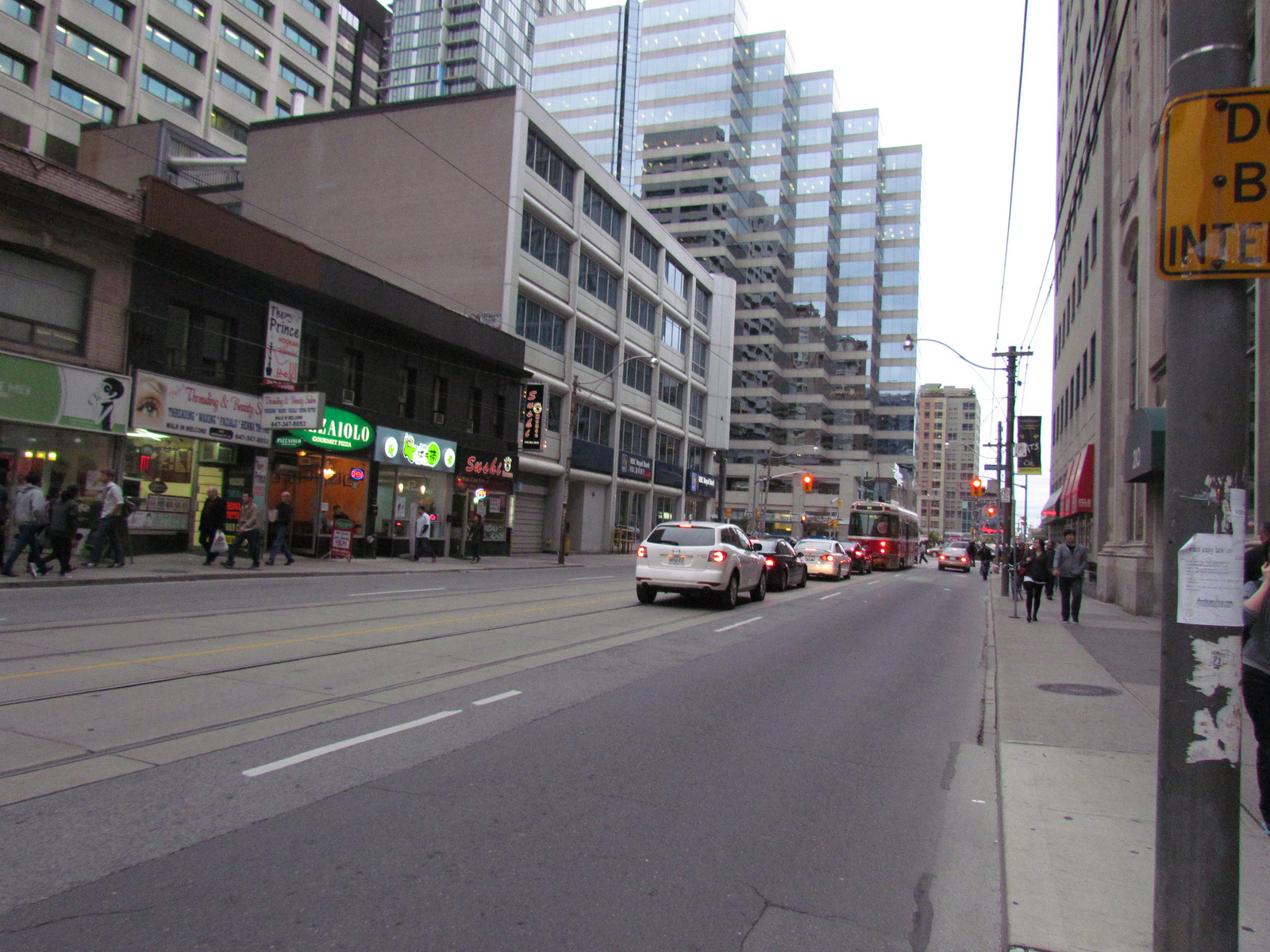} \label{figure:andy}}
\hspace{1pc}
\subfigure[\textit{Celebrating our 6\textsuperscript{th} wedding anniversary in Villa Mary} by Rita \& Tomek \ccbyncsa~\url{https://flic.kr/p/fCXEJi}.]{\includegraphics[width=0.4\textwidth]{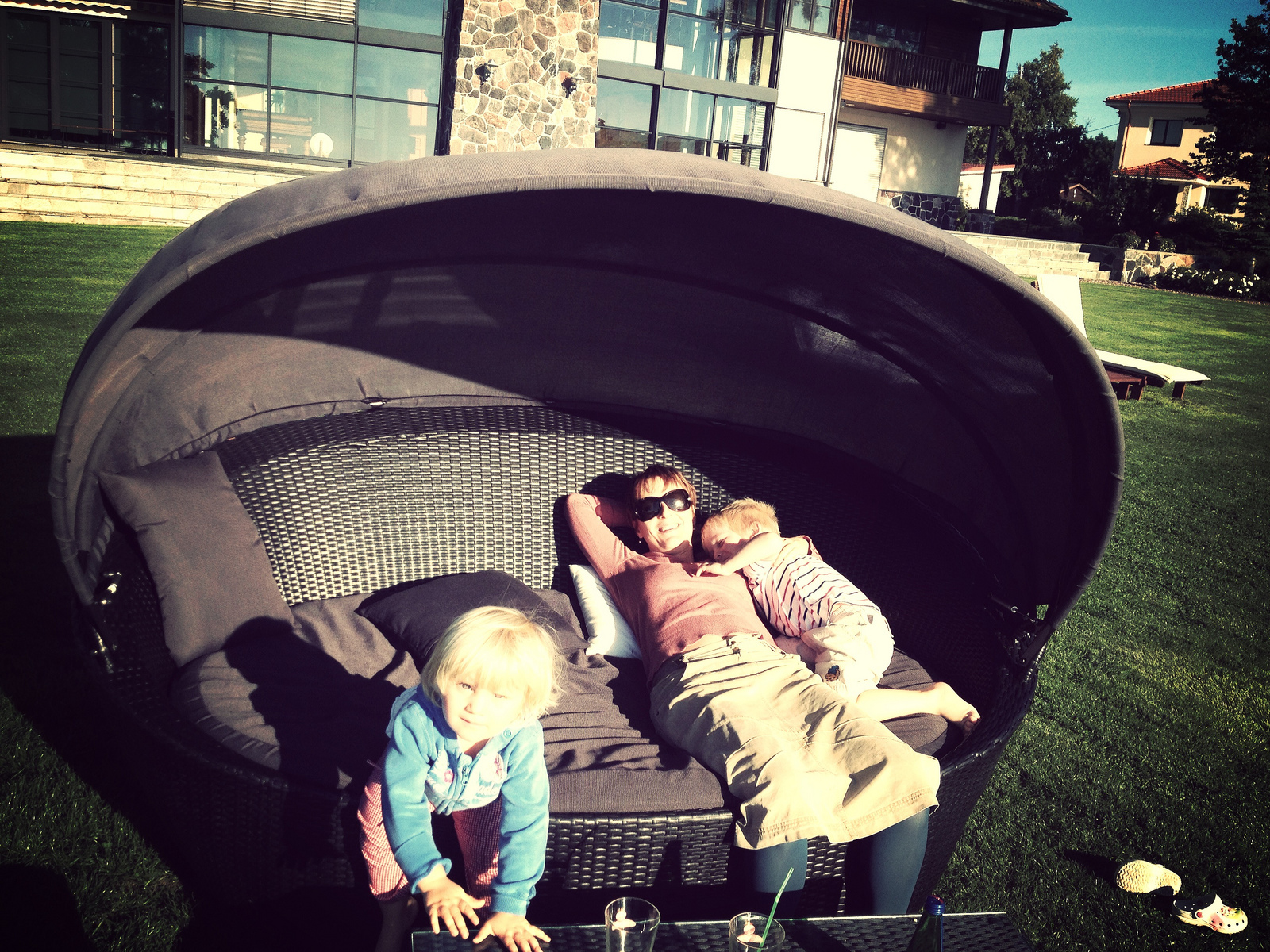} \label{figure:rita}}
\caption{Two photos of real world scenes from photographers in the YFCC100M dataset.}
\label{figure:ccimages}
\vspace{-6pt}
\end{figure*}

\noindent The dataset includes a diverse collection of complex real world scenes, ranging from 200,000 street life-blogged photos by photographer Andy Nystrom, Figure~\ref{figure:andy}, to snapshots of day to day life, holidays and events, Figure~\ref{figure:rita}. To understand more about the visual content represented in the dataset, we used a deep learning approach to find the presence of a variety of concepts, such as people, animals, objects, food, events, architecture, and scenery. Specifically, we applied an off-the-shelf deep convolutional neural network~\cite{Krizhevsky2012} with 7 hidden layers, 5 convolutional layers and 2 fully connected ones. The penultimate layer of the convolutional neural network output was employed as the image feature representation to train the visual concept classifiers. We used Caffe~\cite{Jia:2014cm} to train 1,570 classifiers, each being a binary SVM, using 15 million photos taken from the entire Flickr corpus; positive examples were crowd labeled or handpicked based on targeted search/group results, while negative examples were drawn from a general pool. We tuned the classifiers such that they achieved at least 90\% precision on a held-out test set. In Table~\ref{table:taglist} we show the top 25 detected concepts in both photos and videos (using the first frame). We see a diverse collection of visual concepts being detected, from outdoor to indoor images, sports to art, nature to architecture. As we realize the detected visual concepts may be valuable to the community, we recently released them as one of our expansion packs.

\begin{table}[t]
\centering
\caption{The top 25 of 1,570 visually detected concepts in the YFCC100M dataset. Photos and videos are counted as many times as they contain visual concepts.}
\begin{tabular}{l r r}
\toprule
\textit{Concept} & \textit{Photos} & \textit{Videos} \\
\midrule
outdoor & \texttt{44,290,738} & \texttt{266,441} \\
indoor & \texttt{14,013,888} & \texttt{127,387} \\
people & \texttt{11,326,711} & \texttt{56,664} \\
nature & \texttt{9,905,587} & \texttt{47,703} \\
architecture & \texttt{6,062,789} & \texttt{11,289} \\
landscape & \texttt{5,121,604} & \texttt{28,222} \\
monochrome & \texttt{4,477,368} & \texttt{18,243} \\
sport & \texttt{4,354,325} & \texttt{25,129} \\
building & \texttt{4,174,579} & \texttt{7,693} \\
vehicle & \texttt{3,869,095} & \texttt{13,737} \\
plant & \texttt{3,591,128} & \texttt{11,815} \\
blackandwhite & \texttt{2,585,474} & \texttt{10,351} \\
animal & \texttt{2,317,462} & \texttt{9,236} \\
groupshot & \texttt{2,271,390} & \texttt{4,392} \\
sky & \texttt{2,232,121} & \texttt{11,488} \\
water & \texttt{2,089,110} & \texttt{15,426} \\
text & \texttt{2,074,831} & \texttt{5,623} \\
road & \texttt{1,796,742} & \texttt{12,808} \\
blue & \texttt{1,658,929} & \texttt{10,273} \\
tree & \texttt{1,641,696} & \texttt{6,808} \\
hill & \texttt{1,448,925} & \texttt{6,075} \\
shore & \texttt{1,439,950} & \texttt{8,602} \\
car & \texttt{1,441,876} & \texttt{4,067} \\
head & \texttt{1,386,667} & \texttt{8,984} \\
art & \texttt{1,391,386} & \texttt{2,248} \\
\bottomrule
\end{tabular}
\label{table:taglist}
\end{table}

Flickr makes little distinction between a photo and a video; however, videos do play a role both on Flickr and in this dataset. While photos encode their content primarily through visual means, videos also convey this through audio and motion. Only 5\% of the videos in our dataset do not have an audio track. From a manual examination of over 120 randomly selected geotagged videos having audio, we found that most of their audio tracks were very diverse. Namely, 60\% of the videos were home-video style with little ambient noise; 47\% of the videos had heavy ambient noise such as people chatting in the background, sound of traffic, and wind blowing into microphone; 25\% of the sampled videos contained music, either played in the background of the recorded scene, or inserted at the editing phase; 60\% of the videos did not contain any form of human speech at all, and even for the ones that contained human speech; 64\% were from multiple subjects and crowds in the background speaking to one another, often at the same time. The vocabulary of approximately 280,000 distinct user tags used for the video annotations indeed shows that those describing audio content (\texttt{music}, \texttt{concert}, \texttt{festival}) and motion content (\texttt{timelapse}, \texttt{dance}, \texttt{animation}) are more frequently applied to videos than to photos. When we compare the videos in the dataset to those taken from YouTube in the time period 2007--2012, we note that YouTube videos are on average longer (Flickr: 39 sec., YouTube: 214 sec.). This is likely due to the initial handling of videos on Flickr where their length until recently was restricted to a maximum of 90 seconds; recent videos uploaded to Flickr indeed tend to be longer.

\subsection{Representativeness}
\noindent To create the dataset, we did not perform any specific filtering besides excluding photos and videos that had been marked as `screenshot' or `other' by the Flickr user. We did, however, include as many videos as possible, because videos form a small percentage of media uploaded to Flickr and a random selection would have led to relatively few videos to be selected. We further included as many photos as possible that were associated with a geographic coordinate to encourage spatiotemporal research. Together these photos and videos form approximately half of the dataset, and the remainder is composed of CC photos randomly selected from the entire pool of photos on Flickr.

To investigate whether our dataset contains a representative sample of real world photography, we collected an additional random sample of 100 million public Flickr photos and videos, irrespective of their license, that were uploaded during the same time period as those in our dataset. We then compared the relative frequency with which content and metadata were present in our dataset and in the random sample. We found that the average difference in relative frequencies between two corresponding visual concepts, cameras, timestamps (year and month) and locations (countries) was only 0.02\%, with an overall standard deviation of 0.1\%. The maximum difference we observed was 3.5\%, due to more videos in the YFCC100M having been captured in the United States than in the random sample (46.2\% vs. 42.7\%). While we found little individual differences between the relative frequency of use of any two corresponding cameras in our dataset and in the random sample, at most 0.5\%, we did observe the earlier mentioned tendency towards more professional DSLR cameras in the YFCC100M rather than regular point-and-shoot cameras. This notwithstanding, our dataset appears to overall exhibit similar characteristics to photos and videos present in the entire Flickr corpus.

\subsection{Features and Annotations}
\noindent Computing features for 100 million media items is a time-consuming and computationally expensive task. Not everyone has access to a distributed computing cluster, and just performing light processing of all the photos and videos on a single desktop machine may already take several days. From our experience with organizing benchmarks on image annotation and location estimation we noticed that accompanying the datasets with precomputed features reduced the burden on the participating teams, allowing them to focus on solving the task at hand rather than on processing the data. As mentioned earlier, we are currently computing a wide variety of visual, aural, textual and motion features for the dataset, and already have released several of them. The visual features span the gamut of global (e.g.\ Gist), local (e.g.\ SIFT) and texture (e.g.\ Gabor) descriptors, the aural features include power spectrum (e.g.\ MFCC) and frequency (e.g.\ Kaldi) descriptors, the textual features refer to closed captions extracted from the videos, and the motion features include dense trajectories and shot boundaries. These features, being computed descriptors of the photos and videos, will be licensed without any restrictions under the \texttt{CC0} (\cczero) license. Real world data does not have well-formed annotations, which presents the sensemaking of the dataset itself as an area of investigation. Annotations, such as bounding boxes, segmentations of objects and faces, and image captions are not yet available for the YFCC100M at present, although generating and releasing them is on our roadmap.

\subsection{Ecosystem}
\noindent Our dataset has already given rise to an ecosystem of diverse challenges and benchmarks, similar to how ImageNet, PASCAL and TRECVID have been used by the community. For example, the MediaEval Placing Task~\cite{Choi:2014:PlacingTask}, an annual benchmark in which participants develop algorithms for estimating the geographic location where a photo or video was taken, is currently based on our dataset. To support research in multimedia event detection the YLI-MED corpus~\cite{Bernd:2015ym} was recently introduced, which consists of 50,000 handpicked videos from the YFCC100M that belong to events similar to those defined in the TRECVID MED challenge. Approximately 2,000 videos were categorized as depicting one of ten target events, and 48,000 as belonging to none of these events. Each video was further annotated with additional attributes like language spoken and whether the video has a musical score. The annotations also include degree of annotator agreement and average annotator confidence scores for the event categorization of each video. The authors stated that the main motivation for the creation of the YLI-MED corpus was to provide an open set without the license restrictions imposed on the original TRECVID MED dataset, while it for instance also could serve as additional annotated data to improve the performance of current event detectors.
Other venues in which our dataset currently features are the ACM Multimedia 2015 Grand Challenge on Event Detection and Summarization, as well as the ACM Multimedia 2015 MMCommons Workshop; the latter kicks off the development of a research community around annotating all 100 million photos and videos. Certainly, the utility of the dataset will grow as more features and annotations are produced and shared, whether by ourselves or by others.

\subsection{Strengths and Limitations}
\noindent We specifically note the following strengths ($\oplus$) and limitations ($\ominus$) of the YFCC100M dataset:

\begin{description}[leftmargin=0pt,itemsep=6pt]
\item[$\oplus$ Design.] Our dataset differs in its design from most multimedia collections. The collection of photos, videos, and metadata in our dataset has been curated to be comprehensive and representative of real world photography, expansive and expandable in coverage, free and legal to use, and as such intends to consolidate and supplant many of the existing datasets currently in use. We emphasize that our dataset does not challenge collections that are different and unique (e.g.\ PASCAL, TRECVID, ImageNet, COCO), but we instead aspire to make it the preferred choice for researchers, developers and engineers with small and large multimedia needs that can be easily satisfied by our dataset, rather than having them needlessly collect their own data.
\item[$\oplus$ Equality.] The dataset ensures data equality for research, which facilitates reproducing, verifying and extending scientific experiments and results.
\item[$\oplus$ Volume.] The dataset, spanning 100 million media objects, is the largest public multimedia collection to have ever been released.
\item[$\oplus$ Modalities.] In contrast with most existing collections, our dataset includes both photos and videos, making it a truly multimodal multimedia collection.
\item[$\oplus$ Metadata.] Each media item is represented by a substantial amount of metadata. The dataset includes metadata that is often absent in existing datasets, such as machine tags, geotags, timestamps, and cameras. While social metadata is not included due to their ever-changing nature, it can be easily obtained by querying the Flickr API.
\item[$\oplus$ Licensing.] The vast majority of available datasets include media whose licenses do not allow them to be used without explicit permission from the rights holder. While fair use exceptions may be invoked depending on the nature of use, these will generally not be applicable to research and development performed by industry and/or for commercial gain. For example, a university spin-off offering a mobile product recognition application that displays matching ImageNet images for each detected product will not only violate the ImageNet license agreement, but also very likely copyright law. Our dataset overcomes this issue providing rules on how the dataset should be used to comply with licensing, attribution and copyright.
\item[$\ominus$ Annotations.] Our collection at present reflects the data as it is \textit{in the wild}; there are lots of photos and videos, but they are currently associated with limited metadata and annotations. We note that our dataset may not and/or cannot offer every single kind of content, metadata and annotations present in existing collections (e.g.\  object segmentations as in COCO, broadcast videos as in TRECVID), although we believe that our efforts and those that spring from the ecosystem currently being constructed around the dataset will offer more depth and richness to the existing content, as well as new material, making it more complete and useful over time. While a lack of annotations clearly forms a limitation of our dataset, we also see this as a challenge. At the scale of 100 million media objects there are enough metadata labels for training and prediction of some attributes, e.g.\ geographic coordinates, and there exist opportunities to create new methods for labeling and annotation through explicit crowdsourced work or through more modern techniques using social community behaviors. In addition, given the plurality of existing Flickr datasets and the size of our dataset, some overlaps are to be expected, such that existing annotations directly transfer over to our dataset. Of particular note is COCO of which a third is present (about 100,000 images) in the YFCC100M. Over time we will also release the intersections with known datasets as expansion packs. We envision that the intersection with existing datasets allows researchers to expand upon what is known and actively researched.
\end{description}

\subsection{Guidelines and Recommendations}
\noindent Although we listed volume as a strength of our dataset, at the same time it can also pose a weakness when having insufficient computational power, memory and/or storage available. The compressed metadata of all 100 million media objects requires 12.5GB of hard disk space and, at the default pixel resolution used on Flickr, the photos take up approximately 13.5TB and the videos 3.0TB. While the entire dataset---metadata and/or content---can be processed in a matter of minutes to hours on a distributed computing cluster, it may take a few hours to days to do on a single machine. Naturally, the YFCC100M can still be used for experiments by only focusing on a subset of the data. Also, different fields of research, engineering and science have different data requirements and evaluation needs, and all 100 million media objects in the dataset are not likely to be needed for each and every study. We do note that in the computer science literature it is uncommon for a paper to describe in enough detail how the dataset they used in their evaluations was created, effectively preventing others from faithfully replicating or comparing against the achieved results. One clear challenge for the future is ensuring that subsets of the dataset used in experiments can be accurately reproduced. To this end, we suggest researchers to forego arbitrary selections from our dataset when forming a subset for use in their evaluations, but rather to use a principled approach that can be succinctly described. Such selection logic should examine one or both of the following two aspects of the dataset, namely that (i) the photos and videos in the dataset are already randomized, and (ii) the dataset consists of 10 consecutively numbered files. As such, a selection logic could be as simple as
\begin{quote}
``We used the videos in the first four metadata files for training, those in the following three files for development, and those in the last three for testing.''
\end{quote}
or more complicated as
\begin{quote}
``From all photos taken in the United States, we selected the first 5 million and performed 10-fold cross-validation.''
\end{quote}
Alternatively, the created subset can be made available for download described as a set of object identifiers that index into the dataset. As an example, the organizers of the MediaEval Placing Task made, in addition to the training and test sets, the visual and aural features they extracted from the content available for download. We envision the research community to also follow this way of using and sharing the dataset.

\subsection{Future Directions}
\noindent Our dataset provides opportunities for large-scale unsupervised learning, semi-supervised learning and learning with noisy data. Beyond this, the YFCC100M offers the opportunity to advance research, give rise to new challenges and solve existing ones. For instance, we note the following challenges in which our dataset may play a leading role:

\begin{description}[leftmargin=0pt,itemsep=6pt,topsep=0pt,partopsep=3pt,parsep=0pt,listparindent=\parindent]
\item[AI and Vision.] Large datasets have played a critical role in advancing computer vision research. ImageNet~\cite{Deng:2009jn} itself led the way into the development of advanced machine learning techniques like deep learning~\cite{Krizhevsky2012}. The computer vision as a field now seeks to do visual recognition by learning and benchmarking from large-scale data. The YFCC100M dataset provides new opportunities for this effort by developing new approaches that harness more than just the pixels. For instance, new semantic connections can be made by inferring them through relating groups of visually co-occurring tags in images depicting similar scenes, where such efforts hitherto have been hampered by the lack of a sufficiently large dataset. Our expansion pack containing the detected visual concepts in the photos and videos can assist with addressing this challenge. However, visual recognition goes beyond image classification and involves obtaining a deeper understanding of images. Object localization, object interaction, face recognition, image annotation, are all important cornerstone challenges that will lead the way to retelling the story of an image: what is happening in the image and why was this image taken. Flickr, having a rich diverse collection of image types, provides the groundwork for total scene understanding~\cite{Li2009} in computer vision and artificial intelligence, a crucial task that can be expanded upon with our dataset, and even more so once additional annotations are released by ourselves and by others.
\item[Spatiotemporal Computing.] Beyond pixels, we emphasize time and location information as key components for research that aims at understanding and summarizing events. Starting with location, the geotagged photos and videos, about half of the dataset, provide a comprehensive snapshot of photographic activity over space and time. In the past, geotagged media has been used to address a variety of research endeavors, such as location estimation~\cite{Hays:2008:im2gps}, event detection~\cite{Rattenbury:2007cw}, finding canonical views of places~\cite{Crandall:2009db}, and visual reconstruction of the world~\cite{Snavely:2006bv}. Even understanding styles and habits have been used to reverse lookup authors online~\cite{Hecht:2011:TJB:1978942.1978976}. Geotagged media in the YFCC100M dataset can help push the boundaries in these research areas even further.

More data brings new discoveries and insights, yet it simultaneously also makes searching, organizing and presenting the data and findings more difficult. The cameraphone has enabled people to capture more photos and videos than they can effectively organize. One challenge for the future is therefore devising algorithms that are able to automatically and dynamically create albums for use on personal computers, cloud storage or mobile devices, where desired media and moments of importance can be easily surfaced based on simple queries. Harnessing the spatiotemporal context at capture time and at query time will take a central role in successfully addressing this challenge.

The previous challenge speaks towards social computing efforts aimed at understanding events in unstructured data through the reification of the photo with space and time. While GPS-enabled devices are capable of embedding the precise time, location and orientation of capture in the metadata of a photo, in many instances this information is unavailable or out of date: seconds, hours or sometimes even months. In addition, people frequently forget to adjust the camera clock to the correct timezone when traveling. These kinds of issues pose problems for the accuracy of any kind of spatiotemporal data analysis, and new challenges in computational photography therefore include devising algorithms that can either fix or are resilient against the erroneous.
\item[Digital Culture and Preservation.] What we know to be UGC has grown from simple video uploads and bulletin board systems; life online has come to reflect culture. These large online collections tell a larger story about the world around us, from consumer reviews~\cite{yelp:dataset} that speak to how people engage with the spaces around them to 500 years of scanned book photos and illustrations~\cite{flickr:archive} that describe how concepts and objects have been visually depicted over time. Beyond archived collections, the photostreams of individuals represents many facets of recorded visual information, from remembering moments and storytelling to social communication and self-identity~\cite{dijck2008:digital}. This presents a grand challenge of sensemaking and understanding digital archives from non-homogeneous sources. Photographers and curators alike have contributed to the larger collection of Creative Commons images, yet little is known on how such archives will be navigated and retrieved, or how new information can be discovered therein. Our dataset offers avenues of computer science research in multimedia, information retrieval, and data visualization in addition to the larger questions of digital libraries and preservation.
\end{description}

%%% Local Variables: 
%%% mode: latex
%%% TeX-master: "TMM2015_YFCC100M"
%%% End: 

% !TEX root = CACM2015_YFCC100M.tex

\section{Conclusions}
\noindent Data is one of the core components of research and development. In the field of multimedia, datasets are usually created with a single purpose in mind, and as such lack reusability. Additionally, datasets generally are not or may not be freely shared with others, and as such lack reproducibility, transparency and accountability. To overcome this conundrum, we have released one of the largest datasets ever created, consisting of 100 million media objects published under a Creative Commons license. Our dataset has been curated to be comprehensive and representative of real world photography, expansive and expandable in coverage, free and legal to use, and intends to consolidate and supplant many existing collections.
Our dataset encourages the improvement and validation of research methods, reduces the effort of acquiring data, and stimulates innovation and potential new data uses. We have further provided rules on how the dataset should be used to comply with licensing, attribution and copyright, and offered guidelines on how to maximize compatibility and promote reproducibility of experiments with existing and future work.

%%% Local Variables: 
%%% mode: latex
%%% TeX-master: "TMM2015_YFCC100M"
%%% End: 

% !TEX root = CACM2015_YFCC100M.tex

\section*{Acknowledgements}
\label{section:acknowledgements}
\noindent We would like to thank Jordan Gimbel and Kim Capps-Tanaka at Yahoo, Pierre Garrigues, Simon Osindero, and the rest of the Flickr Vision \& Search team, Carmen Carrano and Roger Pearce at LLNL, as well as Julia Bernd, Jaeyoung Choi, Luke Gottlieb and Adam Janin at ICSI. We are further thankful to ICSI for making their data publicly available in collaboration with Amazon. Portions of this work were performed under the auspices of the U.S. Department of Energy by LLNL under Contract DE-AC52-07NA27344, and supported by the National Science Foundation by ICSI under Award Number 1251276.

%%% Local Variables: 
%%% mode: latex
%%% TeX-master: "TMM2015_YFCC100M"
%%% End: 

\bibliographystyle{acm}
\setlength{\bibsep}{0.5pt plus 0.3ex}
\small{
\bibliography{CACM2015_YFCC100M}
}
\end{document}